**Amorphous Solidification of a Supercooled Liquid in the Limit of Rapid Cooling**

Gang Sun[1] and Peter Harrowell[2]

[1] *Department of Physics, Beijing Normal University, Beijing 100875, China*

[2] *School of Chemistry, University of Sydney, Sydney, New South Wales 2006, Australia*

Abstract

We monitor the transformation of a liquid into an amorphous solid in simulations of a glass forming liquid by measuring the variation of a structural order parameter with either changing temperature or potential energy to establish the influence of the cooling rate on amorphous solidification. We show that the latter representation, unlike the former, exhibits no significant dependence on cooling rate. This independence extends to the limit of instantaneous quenches which we find can accurately reproduce the solidification observed during slow cooling. We conclude that amorphous solidification is an expression of the topography of the energy landscape and present the relevant topographic measures.

# 1. Introduction

Amorphous solids, being non-equilibrium materials, are unfettered by the obligation to adopt the global minimum of the free energy. This freedom, such as it is, is manifest in the dependence of the properties of amorphous solids on the details of their fabrication. In the case of glasses formed by cooling a liquid, both the transition temperature and the mechanical properties of the solid are dependent on the cooling rate. The glass transition temperature $T_g$ has been found [1] to decrease with decreasing cooling rate $\gamma$ according to an empirical relation $T_g(\gamma) = T_0 - \dfrac{B}{\ln(A\gamma)}$. Logarithmic dependencies also appear in the effect of cooling rate on glass properties. The thermal expansion coefficient of silica increases linearly with $\ln(\gamma)$ [2] and the shear modulus of model 2D alloy decreases, roughly, linearly with $\ln(\gamma)$ (at least over the 5 orders of magnitude in cooling rates reported in ref. [3]). These types of results make clear that the magnitude of the cooling rate influences both the nature of the transition from liquid to solid and the properties of the amorphous solid formed.

Recently [4,5], we proposed that the transformation of a supercooled liquid into the amorphous solid on cooling can be usefully described by following the temperature dependence of a suitable structural order parameter – one capable of differentiating the solid from the liquid. How does the cooling rate influence a purely structural account of amorphous solidification? In this paper we examine the cooling rate dependence of amorphous solidification and take advantage of the instantaneous character of the order parameter to extend the range of cooling rates up to the maximum – the instantaneous quench. We compare the influence of cooling rate on both the temperature dependence and energy dependence of the order parameter and establish that the latter shows no dependence on the kinetics of cooling.

To begin, it is useful to establish a general aspect concerning the use of order parameters to describe solid states. Solidity, as measured by the zero frequency shear modulus, is not a property of a single configuration [6]. The infinite frequency or Born modulus can be calculated from a single configuration but does not change significantly in going from the amorphous solid to the liquid (at constant pressure) and so cannot serve as a useful measure of solidity [6]. This problem is simplified, however, if we can include as *a priori* knowledge the fact that the low temperature phase is solid. This is approach is standard in the description of other solidification processes, e.g. crystallization [7]. Armed with this extra information, it is sufficient that we identify an order parameter that clearly differentiates the solid configurations from those of the high temperature liquid. With such an order parameter our structural description of amorphous solidification only has to measure the approach to the solid configurations without having to establish their solidity explicitly.

In this spirit, we introduced an order parameter based on a measure of the capacity of a configuration to restrain the motion of the constituent atoms [4,5]. The restraint order parameter for particle j, $\mu_j$, is defined [5] as

$$\mu_j = \exp\left(-\frac{q^2 <\Delta r_j^2>_{eq}}{6}\right) \qquad (1)$$

where q is the magnitude of the wavevector corresponding to the first peak of the structure factor and

$$<\Delta r_j^2>_{eq} = k_B T \sum_\alpha (\vec{v}_\alpha^j)^2 / \lambda_\alpha \qquad (2)$$

with $\lambda_\alpha$ and $\vec{v}_\alpha^j$, the eigenvalue and jth particle components of the eigenvector of the αth instantaneous normal mode. (The preceding constitutes our definition of the term 'restraint'. As defined, restraint is characteristic of an instantaneous structure and is not, in general, equivalent to the dynamic persistence of a particle about its initial position.) As defined in Eq.2, $<\Delta r_j^2>_{eq}$ is the equilibrium mean squared displacement of particle j for a harmonic Hamiltonian defined by the normal modes. Eq.2 only makes physical sense when all eigenvalues are positive. This is generally not the case with instantaneous normal modes. We resolve this problem [5] by replacing any negative eigenvalue in Eq.2 with a positive effective eigenvalue $\lambda_\alpha^{eff}$ defined as

$$\lambda_\alpha^{eff} = \frac{k_B T}{\int\limits_{-\infty}^{\infty} dA_\alpha A_\alpha^2 \exp\left(-U(A_\alpha)/k_B T\right)} \qquad (3)$$

where U is the true potential energy (i.e. including all anharmonic contributions) and $A_\alpha$ is the amplitude of the $\alpha$th instantaneous normal mode. The integral is carried out keeping $A_\delta = 0$ for all $\delta \neq \alpha$. This recipe for regularizing the unstable modes is not presented as a useful approximation of the short time dynamics (in fact, we have shown that it does not [5]). Our justification for Eq.3 rests on its utility in providing a robust and general method for extracting the restraint parameter, as defined above, from an arbitrary atomic (or molecular) condensed configuration.

## 2. The Influence of Cooling Rate on Amorphous Solidification

The structural passage from liquid to amorphous solid on cooling is quantified in Fig. 1a by plotting <μ>, calculated using Eqs. 1-3, vs temperature for configurations generated using a range of cooling rates, γ. The model used here is the binary Lennard-Jones mixture introduced by Kob & Andersen (KA) [8]. Unless otherwise stated, these calculations were carried out using 5000 particles at fixed NVT (at a reduced density of 1.2) using a particle swap Monte Carlo (MC) algorithm [9] and a potential truncated, shifted and smoothed to ensure that the potential and its first derivative vanish at $r_{cut}$ (equal to 2.5 the appropriate σ). The cooling rate γ is defined as the temperature reduced per MC cycle (where one MC cycle consists of N trial MC moves). The slowest cooling rate used here achieves a similar degree of stability to that recently reported by Parmar, Guiselin and Berthier [10] (see Supplementary Materials for details).

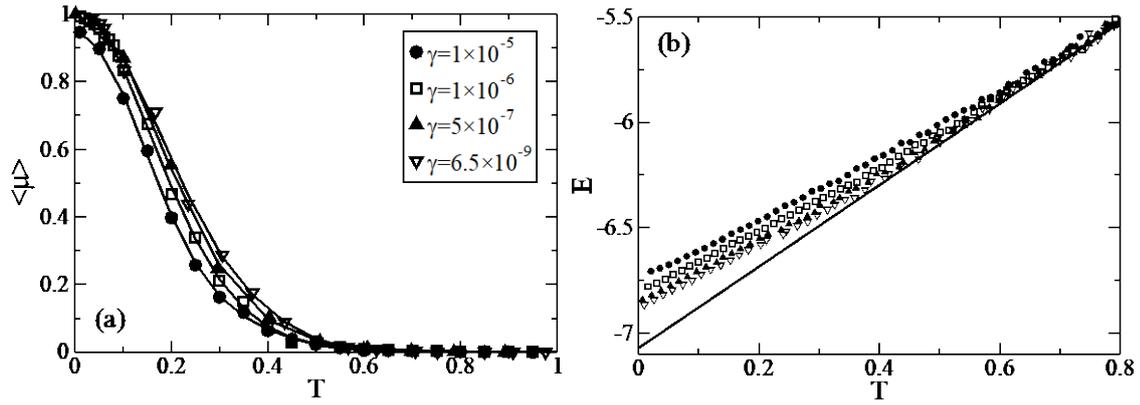

**Figure 1.** a) Plot of <μ> vs T for data from different quench rates γ as indicated. b) Plot of the potential energy E vs temperature for the same set of quench rates. The straight line fit for the slowest quench is used in Eqs. 4 and 5 below to define $T_{eff}$.

Inspection of Fig. 1a shows that the solidification temperature, defined by the point of inflection in the $< \mu >$ vs T curve, shifts to lower values as the cooling rate decreases. While acknowledging that the structural transition described here is quite distinct from the glass transition characterised by dynamics, we note that the observed trend with cooling rate is consistent with previous studies of the influence of the quench rate on potential energy [1], structure [11], shear modulus [3] or yield stress [12]. Close inspection of Fig. 1a also reveals that, for the fastest quench, the restraint does not appear to go to 1 at T = 0 as expected. As shown in the Supplementary Materials, this is a result of a retention, for fast quenches, of unstable modes down to the lowest temperatures studied with an associated decrease in restraint.

The increase in the solidification temperature with decreasing quench rate is similar to the increase in the temperature at which a given potential energy is achieved with decreasing quench rate, as shown in Fig. 1b. This suggests that it is the energy, not the temperature, that provides the better control parameter for the solidification process. To test this idea, we have plotted <μ> as a function of E, the potential energy of the instantaneous configuration, for the runs at different cooling rates (see Fig. 2). The values of <μ> have been averaged over each energy 'bin' for each different value of γ. We find that, in this representation, the dependence on cooling rate has vanished and the curves for different values of γ now lie on top of one another. While the cooling rate may influence the relation between the temperature and the potential energy of the configuration, the restraint is essentially determined by the potential energy alone.

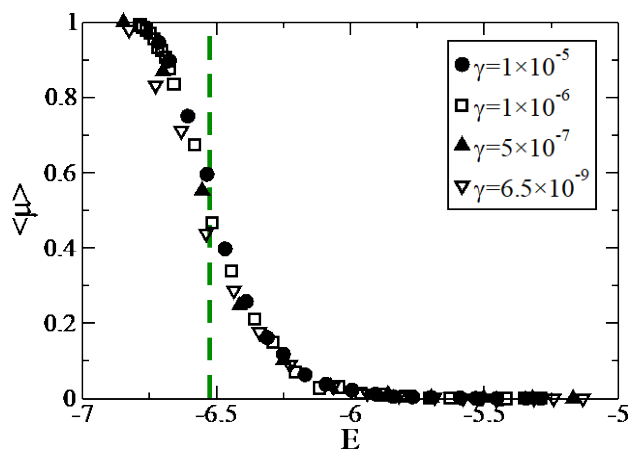

**Figure 2.** Plot of <μ> vs E, the potential energy of the instantaneous configuration, for data from different quench rates used in Fig.1. The characteristic energy, $E_c$, defined in the Supplementary Materials, is indicated by a vertical dashed line.

## 3. Can an Instantaneous Quench Replicate the Solidification at Slow Cooling Rates?

Does this independence of amorphous solidification on cooling rate, when described by the <μ> vs E plot, extend all the way up to maximum possible cooling rate, i.e. an instantaneous quench? To answer this question we have caried out conjugate gradient minimizations starting from a high temperature liquid at T = 1.0, and saved configurations at roughly regular energy intervals during the minimization process. The average restraint is calculated for each configuration using the same approach as above with one change. The temperature that appears in Eqs. 2 and 3 is not well defined during the minimization and so we have used the following ansatz to define an effective temperature, $T_{eff}$. The equilibrated liquid exhibits a linear relation between the average potential energy and the temperature, i.e.

$$\langle E \rangle = E_0 + aT \qquad (4)$$

where, for the KA model, we find $E_0 = -7.07$ and $a = 1.93$ (see Fig. 1b). Using this relation, we can define an effective temperature $T_{eff}$ as,

$$T_{eff} = \frac{(E - E_0)}{a} \qquad (5)$$

where E is the potential energy of the instantaneous configuration. The average restraint from the incremental steps of the minimization, using $T_{eff}$, are plotted in Fig. 3 along with the average restraint from a cooling rate of $\gamma = 6.5 \times 10^{-9}$ from Fig. 1. The agreement between the restraint curves from the slow quench and the instantaneous minimization in Fig. 3 is almost perfect, only faltering at the lowest energies, as shown in the Fig. 3 inset, where, at the same energy, the instantaneous quench exhibits slightly greater restraint than that of the slow quench. (This agreement does not depend on the details of how $T_{eff}$ is obtained from the data in Fig. 1b. See Supplementary Materials.)

Clearly much of amorphous solidification - the transformation of the structural restraint $\langle \mu \rangle$ from 0 to $\sim 0.8$, - is controlled by the potential energy of the configuration alone. How the configuration is generated - whether by slow or instantaneous quench – is irrelevant. The influence of the cooling rate is felt, not in the transformation from liquid to solid, but in the selection of the specific amorphous solid formed. The distinction might be analogous to that between the general description of crystallization and the details of polymorph section.

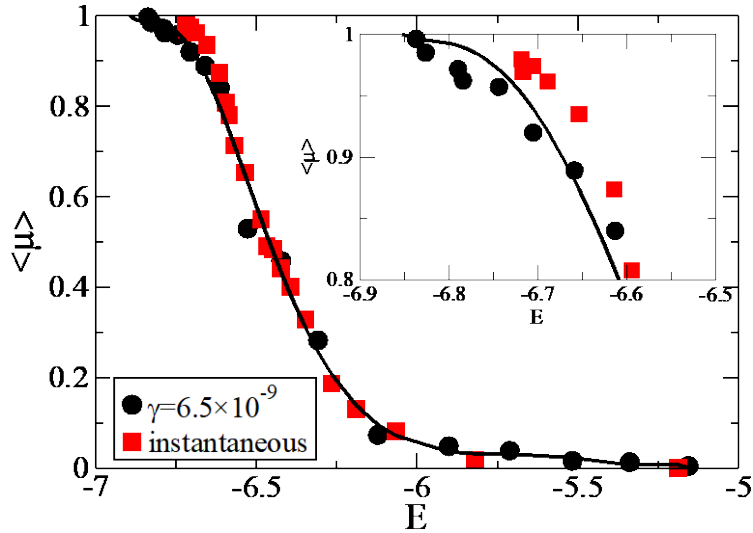

**Figure 3.** The comparison of <μ> vs E for the instantaneous quench (see text) and a slow quench with γ = 6.5 x 10⁻⁹. The curve is a fit to the slow quench data. Inset. A close-up of the variation of <μ> at low energies showing the influence of the cooling rate.

## 4. Quantifying the Change in PEL Topography with Decreasing Energy

Having concluded that amorphous solidification is a direct expression of the topography of the PEL, it remains for us to quantify the specific features of the energy landscape that control the process. The obvious starting point is the fraction of unstable modes, $f_u(E)$, that characterize the energy landscape of instantaneous configurations at different energies.

The decrease in $f_u$ on cooling of a supercooled liquid has been previously reported [13-16]. The assessment of its significance in these previous publications is tangled up with the effort to connect the unstable modes to diffusion, a program that has met with some scepticism [14,16]. The contribution of the unstable modes to the temperature dependence of <μ> can, in contrast, be resolved in a straight-forward fashion by separating their contribution to $< \Delta r_j^2 >_{eq}$, i.e.

$$< \Delta r_j^2 >_{eq} = k_B T \left( \sum_\alpha^{stable} \frac{(\vec{v}_\alpha^{\,j})^2}{\lambda_\alpha} + \sum_\alpha^{unstable} \frac{(\vec{v}_\alpha^{\,j})^2}{\lambda_\alpha^{eff}} \right) \tag{6}$$

$$=< \Delta r_j^2 >_{eq}^{stable} + < \Delta r_j^2 >_{eq}^{unstable}$$

so that, from Eq. 1, we can write

$$\mu_j = \mu_j^{stable} \times \mu_j^{unstable} \tag{7}$$

In Fig. 4 we plot the energy dependence of the system average of the two components of restraint $< \mu^{stable} >$ and $< \mu^{unstable} >$. We find the unstable contribution to the restraint exhibits the larger temperature variation, vanishing at high energies. Since $< \mu > \approx < \mu^{stable} >< \mu^{unstable} >$ (see Supplementary Material), it follows that temperature dependence of $<\mu>$ is dominated by the changing contribution from the unstable modes. This dominance is clearly evident in quantitative similarity between $< \mu^{unstable} >$ and $< \mu >$ in Fig. 4. The contribution of the stable modes, in contrast, exhibits a constant plateau down to an energy of $\sim$ -6.3. It is interesting to note that the magnitude of $< \mu^{stable} >$ is similar to the value of restraint observed in a strong liquid, molten silica, above its melting point [4]. The characteristic energy $E_c$ for solidification coincides with the point at which $< \mu^{unstable} >$ matches this plateau value and, hence, the unstable modes cease to provide any particular reduction in restraint. For energies below $E_c$, the liquid is, structurally, strong [4,5].

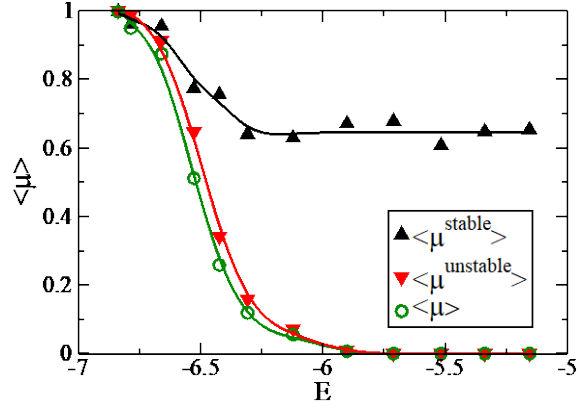

**Figure 4.** The components of restraint, $< \mu^{stable} >$, $< \mu^{unstable} >$ and the total restraint $< \mu >$ as a function of E with the data generated using a slow quench rate of $\gamma = 6.5 \times 10^{-9}$.

To understand the striking energy dependence of $< \mu^{unstable} >$, we can write the contribution of the unstable modes to $< \Delta r_j^2 >_{eq}$ as

$$< \Delta r_j^2 >_{eq}^{unstable} = \frac{E - E_o}{a} (3N-3) f_u(E) g_j(E) \qquad (8)$$

where we have replaced T by the energy equivalent from Eq. 5 and we have introduced $f_u(E)$, the fraction of unstable modes. The quantity $g_j(E)$ is given by

$$g_j(E) = \frac{1}{N_u} \sum_{\alpha}^{unstable} \frac{(\vec{v}_\alpha^j)^2}{\lambda_\alpha^{eff}} \qquad (9)$$

where $N_u$ is the number of unstable modes. $g_j(E)$ corresponds to the average 'bare' contribution (i.e. without the $k_B T$ multiplier) of an unstable mode $\alpha$ to the motion of particle $j$ and, as such, reflects the extent of the participation of the particle motion in the eigenvector of the mode.

As shown the Supplementary Materials, $< g_j(E) >$ exhibits a roughly constant value,

$\overline{g} = 0.0012$, for energies down to $\sim E_c$. This result suggests the following approximation for

the energy dependence of $< \mu^{unstable} >$,

$$< \mu^{unstable} > \approx \exp[-A(E - E_o)f_u(E)] \qquad (10)$$

where $A = \dfrac{q^2}{6a}(3N-3)\overline{g}$. In Fig.5 we find that Eq.10 quantitatively reproduces the main features of the energy dependence of $< \mu^{unstable} >$ using only parameters associated with the energy landscape: $f_u(E)$, $E_o$ and $\overline{g}$, and without any other adjustable fitting parameters. This

means that the striking increase in restraint around the characteristic energy $E_c$ associated

with amorphous solidification (and, with it, the structural origin of fragility [4]) is a

consequence of the increasing rarity of unstable modes with decreasing potential energy. The

success of Eq. 10 also suggests that the increase in localization of the unstable modes, noted

in ref. [17] but neglected in Eq. 10 with the assumption of a constant value for $< g_j(E) >$ (see

Supplementary Materials), does not play a significant role in the solidification process.

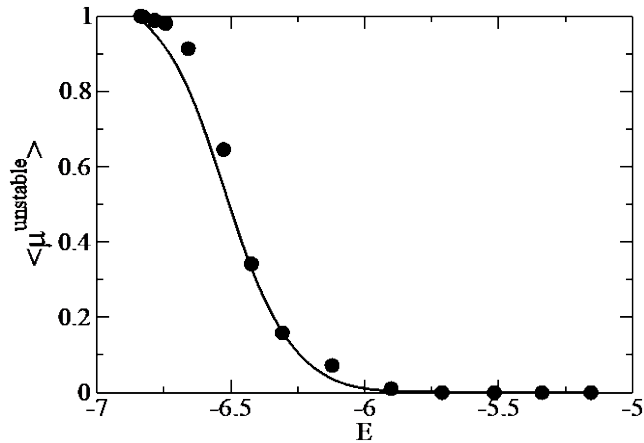

**Figure 5.** The comparison of the energy dependence of $< \mu^{unstable} >$ calculated directly from

simulations (symbols) (i.e, the data from Fig. 3) and the approximate expression from Eq. 10

(solid line).

## 5. Conclusions

We conclude that amorphous solidification of our model glass-former is determined completely by the relation between structural restraint and the topography of the potential energy landscape and that the observed cooling rate dependence is due to the kinetics of energy relaxation that determines the relation between temperature and energy. The proposition that amorphous solidification is a direct expression of the increase in the local curvature of the PEL as the energy decreases is a conceptually simple picture, complicated by the structural variety and spatial heterogeneity of the local minima of the landscape [18]. This conclusion underscores the considerable simplification afforded by treating the transformation of the liquid to the amorphous solid as a structural problem rather than a kinetic one. Our results also clarify what new insights we might expect from advances in our experimental and computational capacity to equilibrate down to lower energies. Improved relaxation of the low temperature liquids is unlikely to shed any new light on the solidification process itself but will provide a deeper understanding of the range of amorphous solids that are possible.

Connections between the disappearance of unstable modes and the glass transition have been previously considered, albeit in a different context to that presented here. In 2002, Grigera et al [19] reported that the index density (i.e. the average number of unstable modes) of the stationary points on the PEL vanished below a threshold value of the potential energy, leaving only local minima at lower energies. This topographical transition in the PEL, described as 'geometric', was correlated with a dynamic crossover. The present work does not address how structure facilitates dynamics but, instead, its compliment, i.e. the capacity of amorphous structure to stabilize themselves. From this perspective, the influence of unstable modes sampled by arbitrary liquid configurations can be clearly established without speculations regarding the trajectories of relaxation at low temperatures.

The challenge remains to connect the properties of the atoms or molecules in the liquid, interaction potentials, shape and flexibility, with the unstable modes of the energy landscape that control the degree of restraint and its associated anisotropy (see ref. [3]). The extension of the restraint analysis to configurations generated by non-thermal routes, as exemplified by our analysis of the instantaneous quench, opens up the prospect of studying the influence of mechanical strain and flow, radiation damage and chemical modification on the structural restraint of the amorphous material. Here it will be interesting to explore the connection between the structural measure used in this paper and measures of local mechanical response [20,21] that have proven successful in characterising heterogeneities in supercooled liquids.

**Acknowledgements**

G. S. gratefully acknowledges the support by the Fundamental Research Funds for the Central Universities.

**Data Availability Statement.** All the data supporting the findings of this study are available within this paper and the Supplementary Information. Additional information is available from the corresponding author upon reasonable request.

**Supplementary Material**

Included in the Supplementary Material is information on the following points:

1. Comparison of quenched energies of this paper and Parmar et al. [1]

2. The definition of $E_c$ in terms of the variance of restraint

3. The Dependence of $<\mu>$ for the Instantaneous Quench on the Details of the Relation between $E$ and $T_{eff}$

4. The Dependence of $f_u$ on T and E for different quench rates.

5. On the relation between $<\mu>$ and the product $<\mu^{stable}><\mu^{unstable}>$

6. The participation ratio of the unstable modes

7. The features of the energy landscape, $f_u(E)$ and $< g_j(E) >$

**SUPPLEMENTARY MATERIAL**

**Amorphous Solidification of a Supercooled Liquid in the Limit of Rapid Cooling**

Gang Sun[1] and Peter Harrowell[2]

*[1] Department of Physics, Beijing Normal University, Beijing 100875, China*

*[2] School of Chemistry, University of Sydney, Sydney, New South Wales 2006, Australia*

<u>Content</u>

**1. Comparison of quenched energies of this paper and Parmar et al. [1]**

The KA model we use in our work is a binary mixture of $N_A$ particles of type A and $N_B$ particles of type B in the ratio $N_A: N_B = 80:20$. The interactions are Lennard-Jones, i.e.

$$\phi_{\alpha\beta}\left(r_{ij}\right) = 4\varepsilon_{\alpha\beta}\left[(\frac{\sigma_{\alpha\beta}}{r_{ij}})^{12} - (\frac{\sigma_{\alpha\beta}}{r_{ij}})^6\right] \qquad (S1)$$

with parameter values: $\varepsilon_{AA} = 1.0, \varepsilon_{AB} = 0.5, \varepsilon_{BB} = 1.5, \sigma_{AA} = 1.0, \sigma_{AB} = 0.8,$ and $\sigma_{BB} = 0.88$. To compute the normal modes of the system, we require that both the energy and the force vanish continuously at the cut-off distance $R_c = 2.5 \, \sigma_{\alpha\beta}$. To meet these conditions, we have used a modified interaction potential, $\tilde{\phi}(r)$, defined as follows,

$$\tilde{\phi}_{\alpha\beta}\left(r_{ij}\right) = \phi_{\alpha\beta}\left(r_{ij}\right) - \phi_{\alpha\beta}\left(R_c\right) - (r_{ij} - R_c)\frac{d\phi_{\alpha\beta}}{dr_{ij}}\bigg|_{r_{ij}=R_c} \quad (S2)$$

This modified potential is clearly different to the bare Lennard-Jones potential. It is also different from the potential used in the MC studies in ref. 1 where the only cut-off condition applied was that the energy vanish at $R_c$, i.e.

$$\phi_{\alpha\beta}^*\left(r_{ij}\right) = \phi_{\alpha\beta}\left(r_{ij}\right) - \phi_{\alpha\beta}\left(R_c\right) \quad (S3)$$

The quantitative differences between the 3 versions of the interaction potential are plotted in Fig. S1 for the case of AA interaction.

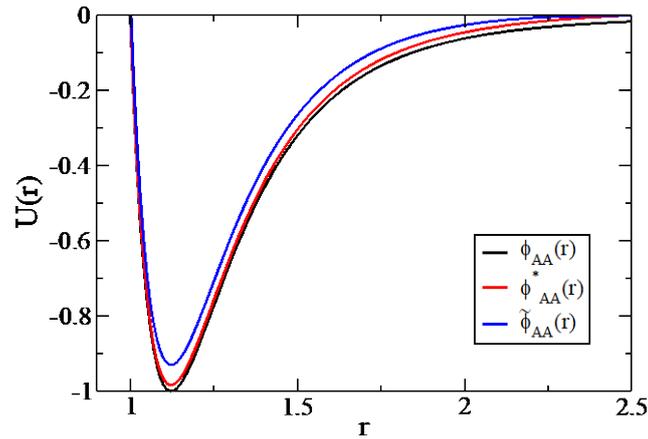

**Figure S1.** Plots of the three versions of the interaction energy between A particles as a function of the separation r. The three versions are: the bare Lennard-Jones potential $\phi_{AA}(r)$ (Eq. S1), the shifted potential used in ref. [1] $\phi_{AA}^*(r)$ (Eq. S3) and the shifted and smoothed potential, $\tilde{\phi}_{AA}(r)$ (Eq. S2), used in this paper.

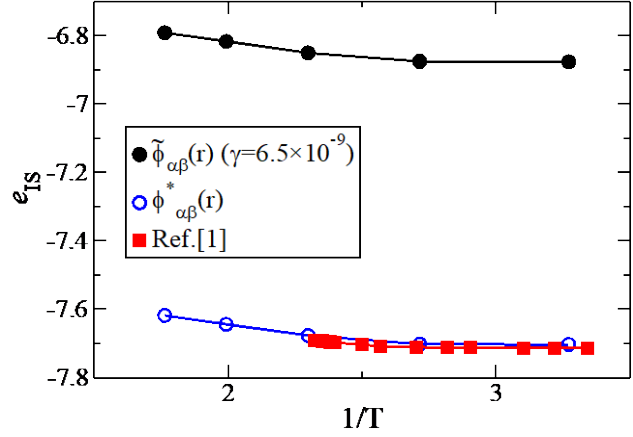

**Figure S2.** Comparison of the potential energy per particle of the local minimum, $e_{IS}$, generated in this work and ref. [1] as a function of the temperature T of the parent liquid. In both cases the liquid was prepared using a slow quench rate $\gamma$ by (standard) MC. The energy from our quench using the shifted and smoothed potential $\tilde{\phi}_{\alpha\beta}(r)$ is plotted as filled circles (black). Using the same configurations, we have recalculated the energy using the shifted potential $\phi^*_{\alpha\beta}(r)$ from ref. [1] and plotted these as open circles (blue). The energies from ref. [1] are plotted as filled squares (red).

As is evident in Fig. S1, $\tilde{\phi}_{AA}(r) > \phi^*_{AA}(r)$ with the difference in energy at $r_{min}$, the position of the minimum, being 0.1. If we wish to compare the energies of quenched configurations from our work with those from ref. [1] we must recalculate the energies of our T = 0 configurations (calculated using $\tilde{\phi}_{AA}(r)$) using the shifted potentials, $\phi^*_{\alpha\beta}(r)$ from ref.[1]. In Fig. S2 we present our values of inherent structure energies obtained by energy minimization from a parent liquid temperature T. The liquid was prepared by an MC simulation using a cooling rate $\gamma = 6.5 \times 10^{-9}$.

With regards the comparison between our work and that of ref. 1 we need to make clear that there are two types of MC results reported in ref. [1]. The focus of this earlier work is what we might called 'augmented swap MC' – 'augmented' by the introduction of a 3$^{rd}$ particle, intermediate in properties between species A and B, that is introduced to increase the probability of acceptance of swap moves, and which is subsequently annealed out of the configurations when the time comes to calculate physical properties. For comparison, the authors also report the results from a 'standard' swap MC which does not use the fictional 3$^{rd}$ particle. The comparisons we present below are between our results and the 'standard' swap MC results from ref. [1], not the 'augmented' results.

The results from ref. [1] are also plotted in Fig. S2. We find that the energies of the configurations generated for this paper are in good agreement with those from ref. [1] after we correct for the different potentials used. The purpose of the comparison is simply to establish that our slow MC quenches have been carried out correctly and are consistent with previous calculations.

## 2. The definition of $E_c$ in terms of the variance of restraint

The characteristic energy $E_c$ referred to in the text is defined as the energy corresponding to the maximum in the variance of the restraint, $< (<\mu>_i - \mu_i)^2 >_i$, where the average $<\dots>_i$ is over the particles in the system. The dependence of the restraint variance on E for different quench rates is plotted in Fig. S3. We find a slight increase in $E_c$ with increasing cooling rate (see Fig.S3).

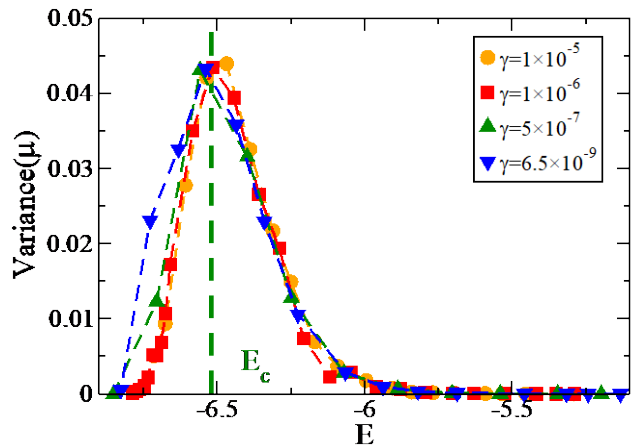

**Figure S3.** The variance of constraint order $\mu$ as function of energy along different cooling rates $\gamma$. The dashed line indicates the critical energy $E_c$, where the variance of $\mu$ exhibits a maximum.

## 3. The Dependence of <μ> for the Instantaneous Quench on the Details of the Relation between E and $T_{eff}$

To check the dependence of the calculation of <μ> for the Instantaneous Quench on the Details of the Relation between E and $T_{eff}$ , we have fitted Eq. 4 to the low energy portion of the data in Fig. 1b to obtain a different set of coefficients (a, $E_0$) – see Fig. S4 – and then recalculated <μ>. The results, shown in Fig. S5, demonstrate that there is no significant difference in <μ> vs E curve.

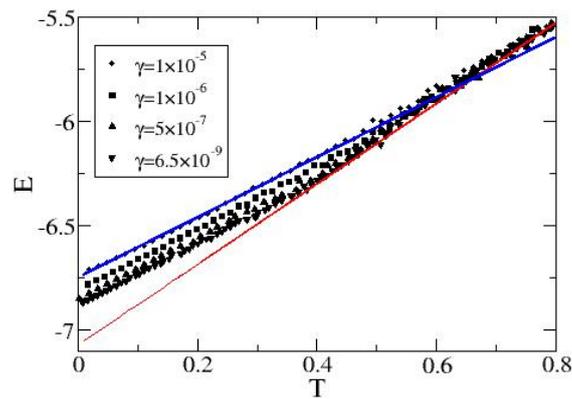

**Figure S3.** Plot of the potential energy E vs temperature for a set of quench rates. Two straight line fits of the form E= aT+$E_0$ are shown: one (red) is to the high energy data and returns a = 1.93 and $E_0$= -7.07 and the second fit (blue) is to the low energy data from the fastest quench and it returns a = 1.41 and $E_0$= -6.75. The fit to high energy is the one used in the text.

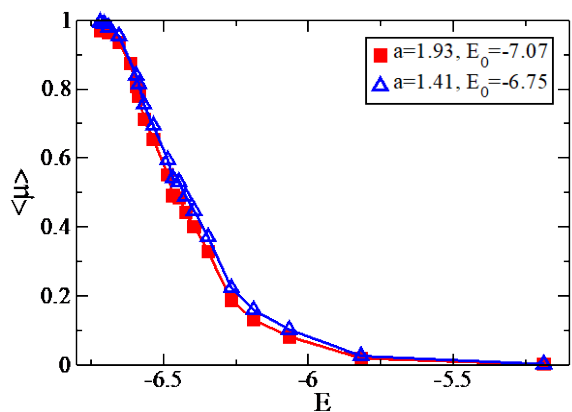

**Figure S5.** Plots of <μ> vs E for the instantaneous quench using the two different sets of coefficients to determine $T_{eff}$. The red points use the coefficients obtained from the fit to the high energy data (as used in the paper) and the blue points use the coefficients from the fit to the low energy data.

## 4. The Dependence of $f_u$ on T and E for different quench rates.

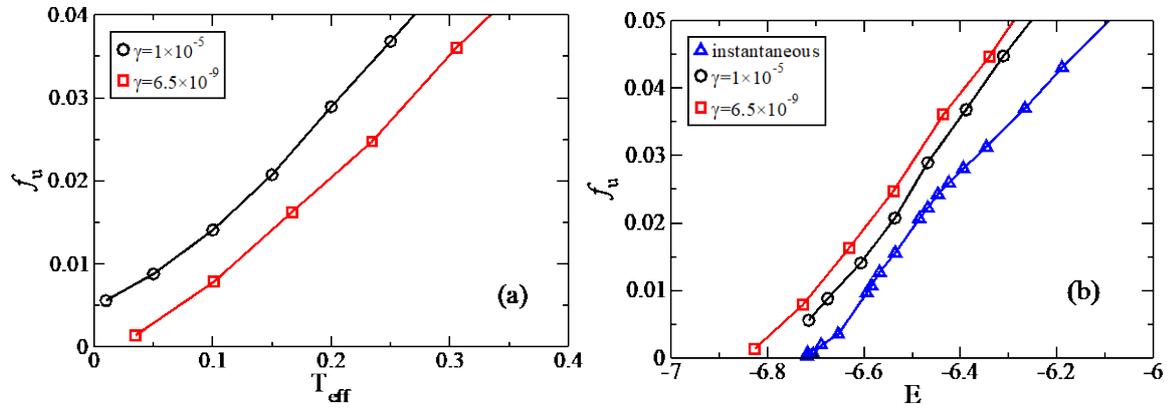

**Figure S6**. The variation of the fraction of unstable modes $f_u$ as a function of a) temperature and b) energy for slow and fast cooling rates, as indicated. The results from the instantaneous quench are included in b).

There are a couple of points worth noting in Fig. S6. First, we find when comparing the slow and fast cooling rates that, for the latter, unstable modes remain present all the way down to the lowest temperatures studied. This is the reason why the associated restraint falls just short of 1 at low T in Fig. 1a; the presence of a handful of modes with very small effective eigenvalues allows for particle motion even at very low temperatures. As shown in Fig. S6b, the variation of $f_u$ with energy shows little variation between large and small values of $\gamma$. The non-vanishing $f_u$ at low T for $\gamma = 10^{-5}$ is the result of the higher energy of this fast quenched system. This leads on to the second point of interest, the difference between the fast quenched systems and that subjected to an instantaneous quench. As shown in Fig.S6b, while both quenches end at similar energies, $f_u$ is systematically lower for the instantaneous quench at a given energy. This result indicates that there is some distribution of $f_u$ for a given value of E and that, within this variability, minimization tends to generate configurations with fewer unstable modes as compared with Monte Carlo sampling.

## 5. On the relation between $<\mu>$ and the product $<\mu^{stable}><\mu^{unstable}>$

In Fig. S7 we confirm numerically that the average value of μ is equal to the product of the average values of $\mu^{stable}$ and $\mu^{unstable}$ as stated in the text.

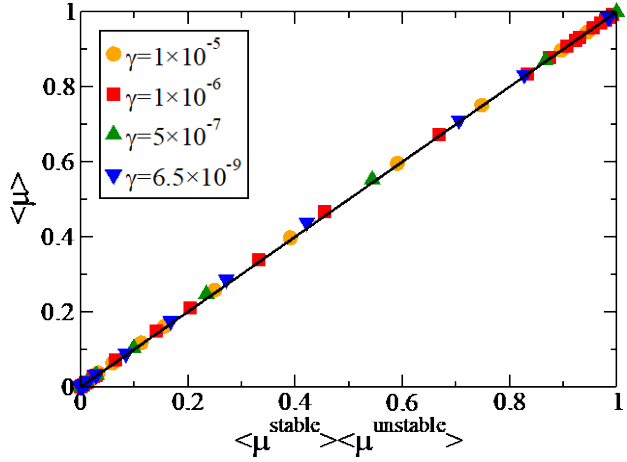

**Figure S7.** The relation between <μ> and the product <μ$^{stable}$><μ$^{unstable}$> on different cooling processes. The black line corresponds to y=x.

The success of this simple factorization suggests that any fluctuations in μ$^{stable}$ and μ$^{unstable}$ are statistically independent.

## 6. The participation ratio of the unstable modes.

Previous work [2] has shown that the unstable modes associated with stationary points of the PEL become more localized as the energy decreases. To see if this trend is also exhibited by unstable modes from general points on the PEL, we have calculated the participation ratio $P_\alpha$ of mode $\alpha$, given by

$$P_\alpha = \frac{\left(\sum_i^N (\vec{v}_\alpha^i)^2\right)^2}{N \sum_j^N (\vec{v}_\alpha^i)^4} \qquad (S4)$$

and the average participation fraction P$_u$ of the unstable modes is $P_u = 1/N_u \sum_\alpha^{unstable} P_\alpha$. In Fig.S8 we plot the energy dependence of P$_u$. Localization, indicated by the decrease in P$_u$, increases steadily below E ∼ -6.3, a result generally consistent with the previous observation [2] for the modes at stationary points. The participation ratio and the number of unstable modes vanish at roughly the same energy.

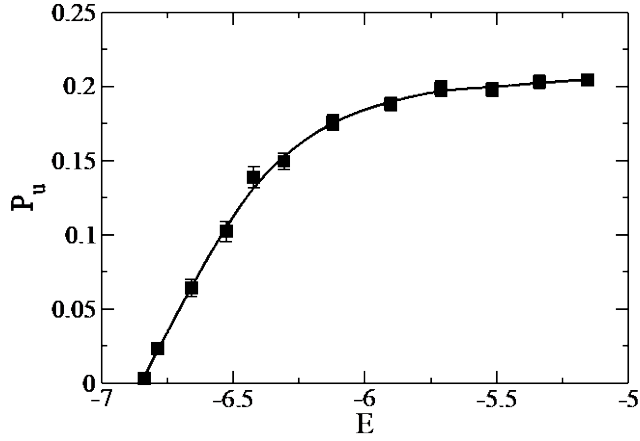

**Figure S68.** A plot of the average participation fraction $P_u$ of the unstable modes as a function of the energy.

## 7. The features of the energy landscape, $f_u(E)$ and $<g_j(E)>$

In Fig. S9 we plot the dependence of $f_u(E)$ and $<g_j(E)>$ (i.e. the average over j) on the energy E. We find that the fraction of unstable modes, $f_u$, decreases monotonically to vanish at an energy of $\sim 6.8$. The quantity $<g_j(E)>$ exhibits a roughly constant plateau at $\bar{g} = 0.0012$ for energies down to $\sim E_c$, below which it decreases in line with the decreasing participation ratio of particles ($P_u$) (see Fig.S8) in the unstable modes. Localization, indicated by the decrease in $P_u$, increases steadily below E $\sim$ -6.3. The participation ratio and the number of unstable modes vanish at roughly the same energy.

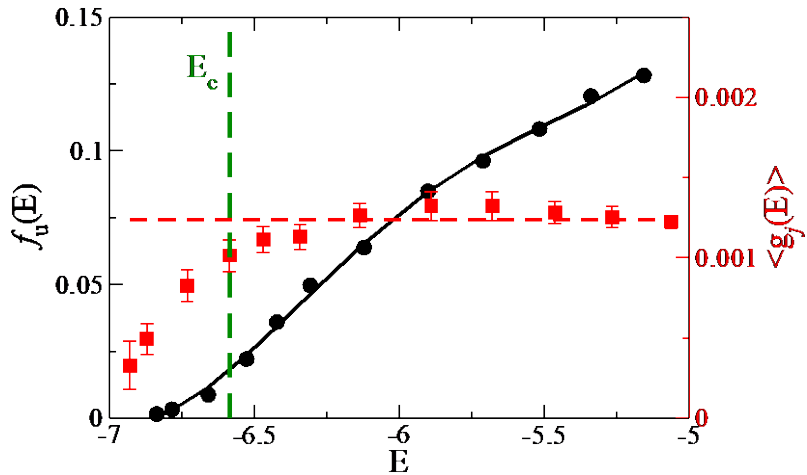

**Figure S9.** The dependence of the fraction of unstable modes $f_u(E)$ and the average 'bare' contribution of unstable modes $<g_j(E)>$ to an individual particle motion as a function of E.

The plateau value of $<g_j(E)>$, $\bar{g} = 0.0012$, is indicated by the horizontal dashed line. The characteristic energy $E_c$ is indicated by the vertical dashed line.

The success of Eq. 10, as demonstrated in Fig. 5 in the paper, can be equally established by comparing the contribution to the mean squared displacement from the unstable modes with the term $(E-E_0)*f_u(E)$. The linear fit demonstrated in Fig. S10 confirms that this simple combination of the fraction of unstable modes and the 'available' energy, $E-E_0$, effectively captures the full energy dependence of the unstable modes' contribution to particle fluctuations.

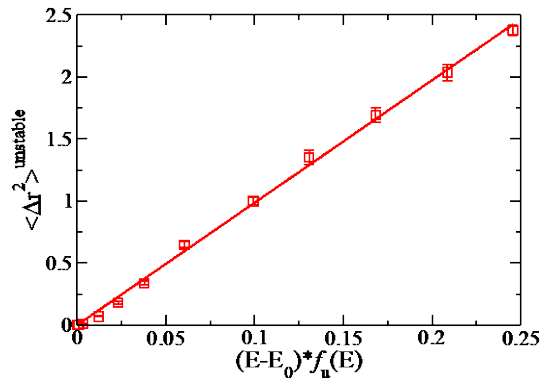

**Figure S10.** The unstable component of mean squared displacement, $<\Delta r^2>^{unstable}$, as function of $(E-E_0)*f_u(E)$. The linear fitting is applied in the figure, the red line.